\documentclass[12pt]{article}
\usepackage{amssymb,amsmath}
\usepackage{epsfig}
\usepackage{epstopdf}
\usepackage{latexsym}
\def\coeff#1#2{\relax{\textstyle {#1 \over #2}}\displaystyle}
\def\ds{\displaystyle}

\def\IR{\mathbb{R}}

\def\cS{{\cal S}}
\newcommand{\be}{\begin{equation}}
\newcommand{\ee}{\end{equation}}
\newcommand{\bea}{\begin{eqnarray}}
\newcommand{\eea}{\end{eqnarray}}

\begin{document}

\begin{titlepage}

\begin{flushright}
\end{flushright}

\bigskip
\bigskip
\centerline{\Large \bf Entropy Enhancement and Black Hole
Microstates}
\bigskip
\bigskip
\bigskip
\centerline{{\bf Iosif Bena$^1$, Nikolay Bobev$^2$, Cl\'{e}ment
Ruef$\, ^1$  and Nicholas P. Warner$^2$}}
\bigskip
\centerline{$^1$ Institut de Physique Th\'eorique, } \centerline{CEA
Saclay, 91191 Gif sur Yvette, France}
\bigskip
\centerline{$^2$ Department of Physics and Astronomy, University of
Southern California,} \centerline{Los Angeles, CA 90089, USA}
\bigskip
\centerline{{\rm iosif.bena@cea.fr\,, ~bobev@usc.edu\,,
~clement.ruef@cea.fr\,, ~warner@usc.edu} }
\bigskip \bigskip

\begin{abstract}

We study fluctuating two-charge supertubes in three-charge
geometries. We show that the entropy of these supertubes is
determined by their locally-defined {\it effective} charges, which
differ from their asymptotic  charges by terms proportional to the
background magnetic fields. When supertubes are placed in deep,
scaling microstate solutions, these effective charges can become
very large, leading to a much larger entropy than one naively would
expect. Since fluctuating supertubes source smooth geometries in
certain duality frames, we propose that such an entropy enhancement
mechanism might lead to a black-hole like entropy coming entirely
from configurations that are smooth and horizonless  in the regime
of parameters where the classical black hole exists.

\end{abstract}

\end{titlepage}

\section{Introduction}

There is a significant body of evidence that supports the idea that,
within string theory, one can resolve BPS black hole singularities
in terms of regular, horizonless {\it microstate geometries}.  These
geometries describe the microstates of black holes in the same
regime of parameters where the classical black hole exists (see
\cite{review, Bena:2007kg, kostas} for reviews). One of the primary
issues in proving this idea is whether the known microstate
geometries represent typical black hole microstates or whether they
are somehow confined to a peculiar atypical sub-sector of the
Hilbert space.

To refine this issue, one should first note that in the large-$N$
limit,  bulk classical geometries describe, to arbitrary accuracy,
bulk quantum states that are dual to coherent states within the
Hilbert space of states of the dual CFT.    Coherent states can
always be used to provide a basis for the Hilbert space, but this
may not be so for the ``semi-classical states'' described by
classical geometries.  Indeed, as one finds with two-charge
geometries, some of the boundary coherent states in such a basis
will be dual to geometries that have string-scale features and for
which the supergravity approximation breaks down or is, at best, a
heuristic guide.

These issues are, however, not directly relevant if one's goal is to
argue that the entropy of a black hole comes from horizon-sized,
horizonless configurations that have unitary scattering and hence no
information loss:   For this, the relevant question is whether the
states corresponding to such smooth microstate geometries are
suitably dense within the Hilbert space of states. Indeed, in the
vicinity a single, smooth microstate geometry that is well-described
in supergravity there might exist a vast (but controlled) number of
quantum microstates that have the same essential features (such as
size, absence of horizon and sub-leading dipole fields). Thus the
classical microstate geometry would act as a representative of these
quantum microstates.

Hence, in counting semi-classical microstate geometries the first
goal is  to  get   the correct dependence of the number geometries
as a function of the charges.  For BPS black holes in five
dimensions, this means one must have:
\begin{equation}
S  ~\sim~\log(N)  ~\sim~    \sqrt{Q_1 \, Q_2 \,Q_3}  \,.
 \label{Sgrowth}
\end{equation}
If  $N_{quantum}$ and $N_{geom}$   respectively represent the number
of quantum states and the number of semi-classical, microstate
geometries that are valid in the supergravity approximation, then
one can recover (\ref{Sgrowth}), if $\log(N_{geom})$ and
$\log(N_{quantum})$ have the same growth to leading order in the
charges\footnote{In this sense, ``suitably dense'' can, in fact,
amount to  an extremely sparse relative population.}.

A subsequent goal is   to get the correct coefficient, $S  =2 \pi
\sqrt{Q_1 Q_2 Q_3} $, which amounts to predicting the correct
central charge for the underlying conformal field theory.  On the
other hand, if one restricts oneself to a finite fraction of the
degrees of freedom (with, perhaps, a lower central charge) and
obtains   horizon-sized, horizonless black-hole microstates with
unitary scattering, it seems very implausible that restoring the
rest of the degrees of freedom will drastically change the
macroscopic features of these microstates.  In particular, it is
very unlikely that restoring such degrees of freedom will generate a
horizon.

Thus, establishing that black holes in string theory are ensembles
of horizonless configurations with unitary scattering is not as
demanding as it might, at first, seem, and could reduce to showing
that a semi-classical counting of smooth, horizonless, classical
microstate geometries gives a  black-hole-like, or macroscopic,
entropy (\ref{Sgrowth}). Indeed, it is our purpose here to display a
mechanism by which smooth microstates of such a large entropy can
arise.

In \cite{Bena:2006kb, Bena:2007qc} it was argued that the deep, or
scaling, microstate geometries are the gravitational duals of
states that belong to the ``typical sector'' of the D1-D5-P  CFT.
This was based upon the fact that a typical excitation of the
gravitational system had precisely the correct energy to be the dual
of an excitation in the sector of the CFT that contributes maximally
to the entropy.  In particular, the gravitational red-shift of the
throat provides a critical factor in arriving at the proper
excitation energies. Thus deep, or scaling geometries
\cite{Bena:2006kb, Bena:2007qc, Bates:2003vx} will be one of the
crucial ingredients in accounting for the entropy of black holes
using microstate geometries.

Another important ingredient in our discussion will be the fact that
two-charge supertubes \cite{Mateos:2001qs}, which can have arbitrary
shapes, give smooth supergravity solutions in the duality frame in
which they have D1 and D5 charges \cite{Lunin:2001jy,Lunin:2002iz}.
This has been very useful in matching the entropy of two-charge
smooth supergravity solutions to that of the dual CFT
and served as one of the motivations of the formulation of the
fuzzball proposal. However, even if supertubes can have arbitrary
shapes, and hence a lot of entropy, their naive quantization cannot
hope to account for the entropy of a black hole with a non-trivial,
macroscopic horizon (\ref{Sgrowth}). Indeed, as found in
\cite{Palmer:2004gu,Bak:2004kz,Rychkov:2005ji}, since supertubes
only carry two charges, their entropy scales like:
\begin{equation}
S  ~\sim~    \sqrt{Q_1 \, Q_2 }  \,.
 \label{Stube}
\end{equation}
The new insight here comes from considering supertubes in the
background of a scaling geometry. We generalize the analysis of
\cite{Palmer:2004gu}, and use the supertube DBI--WZ action to count
states of quantized supertubes in non-trivial background geometries.
We find that, for the purposes of entropy counting, the supertube
charges $Q_I$ that appear in (\ref{Stube}) are replaced by the local
effective charges of the supertube, $Q_I^{eff}$, which are a
combination of the supertube charges and terms coming from the
interaction between the supertube magnetic dipole moment and the
background magnetic dipole fields.

If there are strong magnetic fluxes in the background, as there are
in a deep, bubbled microstate geometries, these effective charges
can be {\it much larger} than the asymptotic charges of the
configuration, and can thus lead to a very large entropy
enhancement!  Indeed, one finds that if the supertube is put in
certain deep scaling solutions, the effective charges can diverge if
the supertube is suitably localized or if the length of the throat
goes to infinity.  Of course, this divergence is merely the result
of not considering the back-reaction of the wiggly supertube on its
background: Once this back-reaction is taken into account, the
supertube will delocalize and the fine balance needed to create
extremely deep scaling solutions might be destroyed if the tube
wiggles too much.

Hence, we expect a huge range of possibilities in the the
semi-classical configuration space, from very shallow solutions to
very deep solutions.  In very shallow solutions, the supertubes can
oscillate a lot, but they will not have their entropy enhanced and
for very deep solutions the supertube will have vastly enhanced
charges but, if the solution is to remain deep, the supertube will
be very limited in its oscillations.  One can thus imagine that the
solutions with most of the entropy will be intermediate, neither too
shallow (so as to obtain effective charge enhancement), nor too deep
(to alow the supertube to fluctuate significantly).  To fully
support this intuition one will need to construct the full
back-reacted supergravity solution for wiggly supertubes in bubbling
three-charge backgrounds.  Even though we do not yet have such
solutions, it is possible to use the $AdS$-CFT correspondence to
estimate the depth of the bulk microstate solutions dual to states
in the typical sector of the dual CFT \cite{Bena:2006kb}.  We will
use this result to determine the depth of the typical throat and
then argue that the effective supertube charges corresponding to
this throat depth yields an enhanced supertube entropy that is
macroscopic (\ref{Sgrowth}).

It is interesting to note that entropy enhancement is not just a
red-shift effect: There is no entropy enhancement unless there are
strong background magnetic fluxes.  A three-charge BPS black hole
will not enhance the entropy of supertubes: it is only solutions
that have dipole charges, like bubbled black holes or black rings
that can generate supertube entropy enhancement.

The last ingredient that we use is the generalized spectral flow
transformation \cite{Flowpaper}\footnote{See
\cite{Giusto:2004kj,Ford:2006yb} for relevant earlier work.} that
enables us to start from a simple, bubbled black hole microstate
geometry \cite{Bena:2005va,Berglund:2005vb} and generate a bubbled
geometry in which one or several of the Gibbons-Hawking (GH) centers
are transformed into smooth two-charge supertubes. Indeed, from a
six-dimensional perspective (in a IIB duality frame in which the
solution has D1-D5-P charges) this mapping is simply a coordinate
transformation.  One can then study the particular class of
fluctuating microstate geometries that result from allowing the
supertube component to oscillate in the deep bubbled geometries.
The naive expectation is that one would recover an entropy of the
form (\ref{Stube}) but, as we indicated, the $Q_I$ are replaced by
the enhanced $Q_I^{eff}$, and the entropy of these supertubes can
become ``macroscopic'' in that it corresponds to the entropy of a
black hole with a macroscopic horizon. One can then undo the
spectral flow to argue that this entropy is present in the BPS
fluctuations of three-charge bubbling solutions in {\it any} duality
frame.  In fact, spectrally flowing configurations with oscillating
supertubes into other duality frames is not strictly speaking
necessary for the purpose of illustrating entropy enhancement and
arguing that smooth solutions can give macroscopically large
entropy. After all, one could do the full analysis in the D1-D5-P
duality frame and consider smooth black hole microstates containing
both GH centers and supertubes. Nevertheless, since such solutions
have not been studied in the past in great detail, it is easiest to
construct them by spectrally flowing multi-center GH solutions,
which have been studied much more and are better understood.

The fluctuations we consider do not represent the most general,
regular fluctuation of the geometry, but as we outlined earlier,
this is not the point: They represent a sub-sector of the possible
fluctuations whose Hilbert space has entropy that grows much faster
than (\ref{Stube}) and indeed might grow as fast as (\ref{Sgrowth}).
Thus we believe that these microstate geometries may be capturing
generic states of the CFT for black holes and black rings with
non-zero horizon area and capturing enough of them to account for
that horizon area, up to overall numerical factors.  The fact that
we are only looking at a special class of fluctuations means that we
are necessarily restricting the degrees of freedom of the
fluctuations and so one would not expect, at the first pass, to
recover the correct numerical factors in (\ref{Sgrowth}).  The
important progress here is that we see how microstate geometries may
indeed capture enough entropy to account for macroscopic horizons
and for their dependence upon charges.

It is also interesting to note that a similar conclusion -- that
deep, scaling, horizonless configurations can give a macroscopic
(black-hole-like) entropy -- was also reached in \cite{Denef:2007yt}
and \cite{Denef:2007vg}. In \cite{Denef:2007yt} this was done by
considering D0 branes in a background of D6 branes with world-volume
fluxes, in the regime of parameters where the D0 branes do not
back-react.  In \cite{Denef:2007vg}, a similar result was obtained
by studying the quiver quantum mechanics of multiple D6 branes, in
the regime where the branes do not back-react, but form a
finite-sized configuration. Since these computations were performed
in a regime in which the gravitational back-reaction of all or some
of the branes is neglected, it is not clear how the configurations
that give the black hole entropy will develop in the regime of
parameters in which the classical black hole exists, and {\it all}
the branes back-react on the geometry. Their size will  continue
increasing at the same rate as the would-be black hole horizon, and
since they are made from primitive branes, it is very unlikely they
will develop a horizon.  Hence these two calculations do suggest
that the black hole entropy comes from horizonless configurations.
However, since the D0 branes give rise to naked singularities, the
naive strong-coupling extrapolation of these microstate
configurations will not be reliable when the classical black hole
exists.

The microstates that we consider here are also counted in a regime
of parameters in which some of their components, {\it i.e.} the
supertubes, are treated as probes and described by their DBI--WZ
action, and hence do not back-react on the geometry.  However,
unlike the configurations mentioned above, we understand very well
what the supertubes become in the regime of parameters where the
black hole exists: They give rise to smooth horizonless microstate
solutions. Indeed, as we will show in \cite{bigpaper}, the DBI--WZ
description of supertubes gives configurations that in the D1-D5-P
duality frame are smooth in supergravity. Hence, our entropy
calculation is expected to extend to the regime of parameters where
the classical black hole exists.

In section 2 we review the form of general BPS supergravity
solutions with a Gibbons-Hawking (GH) base and in section 3 we
compute the entropy of two-charge supertubes in such solutions. In
Section 4 we discuss the entropy enhancement mechanism, and in
section 5 we consider how the effective charges that give the
enhanced entropy behave in deep scaling solutions. Section 6
contains conclusions.

\section{Fluctuating supertubes in non-trivial backgrounds}

Three-charge bubbling solutions that have the same charges and
dipole moments as black holes and black rings are determined by
specifying a four-dimensional base space, and solving a set of
linear equations to determine the warp factors, and the other
parameters of the solution \cite{onering}.

In the duality frame where the charges of the solutions correspond
to D0 branes, D4 branes and F1 strings, the metric and the dilaton
have the form:
\begin{eqnarray}
ds^2_{10} &=& - \frac{1}{Z_3\sqrt{Z_1Z_2}}\,(dt+k)^2 +
\sqrt{Z_1Z_2}\, ds^2_4 +\frac{\sqrt{Z_1Z_2}}{Z_3}\, dx_5^2 +
  \sqrt{\frac{Z_1}{Z_2}}\,(dx_1^2+dx_2^2+dx_3^2+dx_4^2) \,, \nonumber \\
\Phi &=& \frac{1}{4} \log \left( \frac{Z_1^3}{Z_2Z_3^2}\right),
\end{eqnarray}
where we parameterize the $S^1$ of the F1 string by $x^5$, and the
$T_4$ of the D4 world-volume by $x_i$, $i=1,\dots,4$.  The warp
factors $ Z_1, Z_2, Z_3$ correspond to D0, D4 and F1 charges
respectively.

When the four-dimensional base of the solution is a multi-center
Gibbons-Hawking (Taub-NUT) space,
\begin{equation}
ds_4^2 ~=~ V^{-1} \, \big( d\psi + A)^2  ~+~ V\, d \vec y \cdot d
\vec y \,, \label{GHmetric}
\end{equation}
the full solution can be determined in terms of eight harmonic
functions, $V,K^{I},L_{I},M$ ($I=1,2,3$) on the $\mathbb{R}^3$
spanned by $(y_1,y_2,y_3)$.  As shown in \cite{bigpaper}, the RR
potentials are given by
\begin{equation}
C^{(1)} = (Z_1^{-1} - 1) dt + Z_1^{-1} k - \zeta^{(1)} \,,
\end{equation}
\begin{equation}
C^{(3)} = - Z_3^{-1} (dt + k)\wedge\zeta^{(1)}\wedge dx_5 - dt\wedge
A^{(3)}\wedge dx_5 + (\nu_a + V^{-1}K^{3} \xi_a^{(1)})
\Omega_{-}^{(a)} \wedge dx_5 \,,
\end{equation}
where
\begin{equation}
\vec{\nabla} \times \vec{\nu} = - \vec{\nabla} L_2 \qquad \text{and}
\qquad \Omega_{-}^{(a)} = (d\psi + A)\wedge dy^a - \frac{1}{2}
\epsilon_{abc} dy^b\wedge dy^c \,.
\end{equation}
\begin{equation}
\zeta^{(I)}  ~=~  V^{-1} K^{I} \, (d \psi + A) ~+~ \xi^{I} \,,
\qquad \vec \nabla \times \vec \xi^{I}  ~\equiv~ - \vec \nabla K^I
\,. \label{GHdipolepotentials}
\end{equation}
The functions, $Z_I$, and the angular momentum one-form, $k$, are
\begin{equation}
Z_I  ~=~  \coeff{1}{2} \,V^{-1}\,  C_{IJK} K^{J}K^{K} ~+~ L_I  \,,
\qquad k ~=~ \mu (d\psi+ A) ~+~ \omega  \,, \label{GHfunctions}
\end{equation}
where
\begin{equation}
\mu ~=~ \coeff{1}{6} \,V^{-2}\, C_{IJK}\, K^{I}K^{J}K^{K}  ~+~
 \coeff{1}{2} \,V^{-1}\,   K^{I}L_{I} ~+~ M \,.
\label{GHmu}
\end{equation}
and the one form, $\omega = \vec{\omega}\cdot d\vec{x}$, is given by the
solution of the equation
\begin{equation}
\vec{\nabla}\times \vec{\omega} ~=~ V\vec{\nabla}M - M\vec{\nabla}V
~+~ \coeff{1}{2} \big( K^{I}\vec{\nabla}L_{I} -
L_{I}\vec{\nabla}K^{I}\big)~. \label{GHomegaeqn}
\end{equation}
We take the harmonic functions to have the form:
\begin{eqnarray}
V &~=~& \epsilon_0 ~+~ \sum_{j=1}^{N} \frac{q_j}{r_{j}}~,
\qquad\qquad K^{I} ~=~ \kappa_0^{I} ~+~  \sum_{j=1}^{N}
 \frac{k_j^{I}}{r_{j}}~, \\
L_{I} &~=~&  l_0^{I} ~+~ \sum_{j=1}^{N} \frac{l_j^{I}}{r_{j}}~,
\qquad\qquad M ~=~ m_0 ~+~ \sum_{j=1}^{N} \frac{m_j}{r_{j}} \,,
\label{GHharmonicfn}
\end{eqnarray}
where $r_{j} = |\vec{y}-\vec{y}_j|$, for $N$   Gibbons-Hawking (GH)
points located at  $\vec{y}_j$. To ensure that the solution is
regular (up to $\mathbb{Z}_{q_j}$ orbifold singularities) at $r_{j}
\to 0$ we must have $q_{j}\in \mathbb{Z}$ and
\begin{equation}
l_{j}^{I} ~=~  - \coeff{1}{2} \,q_j^{-1}\,  C_{IJK}
k_{j}^{J}k_{j}^{K} \,, \qquad m_{j} = \coeff{1}{2}  \,q_j^{-2}\,
k_{j}^{1}k_{j}^{2}k_{j}^{3} \,, \qquad j = 1, \ldots, N \,.
\label{GHregular}
\end{equation}

As shown in \cite{Flowpaper}, the spectral flow transformation:
\begin{eqnarray}
\widetilde L_I &~=~&   L_I ~-~  2\, \gamma_I\, M \,, \qquad
\widetilde M ~=~ M \,, \qquad \vec{\tilde \omega} ~=~ \vec \omega  \,, \\
\widetilde K^I &~=~&  K^I ~-~
C^{IJK}\,\gamma_J \,L_K ~+~  C^{IJK}\, \gamma_J\, \gamma_K\, M \,,\\
\widetilde V &~=~& V +  \gamma_I \, K^I - \coeff{1}{2}\, C^{IJK} V
\, \gamma_I \, \gamma_J \, L_K  + \coeff{1}{3}\,  C^{IJK} \gamma_I\,
\gamma_J\, \gamma_K \, M \,,
\end{eqnarray}
transforms solution to solutions, and can change a GH centers into
 other GH centers, or two-charge supertubes\footnote{From a
 four-dimensional perspective this corresponds to transforming a
 primitive D6 brane into a primitive D4 brane.}. This can be arranged
 to happen at the $N^{\rm th}$ GH point by choosing:
\begin{equation}
\gamma_{1,3} =0~, \qquad\qquad \gamma_2= \gamma  ~=~ -
\frac{q_N}{k_N^2}\,, \label{Stubeflow}
\end{equation}
which induces the following changes:
\begin{eqnarray}
\widetilde Z_1 &~=~&   { V   \over \widetilde V} \, Z_1 \,, \qquad
\widetilde Z_3 ~=~    { V   \over \widetilde V} \, Z_3 \,, \qquad
 \widetilde Z_2 ~=~  {\widetilde  V   \over V} \, Z_2 ~-~ 2 \gamma \mu ~+~
  \gamma^2 \, { Z_1\,Z_3  \over \widetilde  V} \,,  \nonumber \\
  \tilde \mu &~=~&   { V   \over \widetilde V} \, \bigg( \mu ~-~   \gamma
\,  { Z_1\,Z_3 \over   \widetilde  V} \bigg)  \,, \qquad \widetilde
V ~=~ V ~+~ \gamma\, K^2  \,. \label{tildefns}
\end{eqnarray}
As explained in \cite{Flowpaper}, the dipole charge and ``bare''
electric charges of the resulting supertube are given by the
coefficients of the divergent terms in $\widetilde K^2$, $\widetilde
L_1$ and $\widetilde L_3$. We define the ``effective'' charges of
the supertube by the divergence of the electric potentials, $Z_I$,
near the supertube:
\begin{equation}
Q_1^{eff} ~\equiv~ 4 \, \lim_{r_N \to 0} \,  r_N \,\widetilde Z_1
~=~ 4 \,q_N \,  \big( \widetilde V^{-1} \, Z_1\big)\big|_{r_N =0}
~=~ 4 \tilde \ell_N^1 ~+~ 4 \tilde k_N^2 (\widetilde{V}^{-1}
\widetilde{K}^3)|_{r_N=0}\,,
 \label{QIeff}
\end{equation}
and similarly for $Q_3^{eff}$. As we will see later, these are the
charges that determine the entropy of supertubes, and since
$(\widetilde{V}^{-1} \widetilde{K}^3) $ depends critically on the
position of the supertube, the effective charges can be much larger
than the asymptotic charges of the system.  This is the crucial
ingredient of the entropy enhancement mechanism.

\section{The probe calculation}

Consider a probe supertube with D0 and F1 charges and D2 dipole
charge in the
three-charge solution with a Gibbons-Hawking base described above. We choose the
supertube world-volume coordinates $\xi$ to be $(t,\theta = \psi ,
z=x^5)$, where $\psi$ is the $U(1)$ fiber of the GH base.

The  DBI--WZ  action of the supertube is:
\begin{multline}
S = T_{D2} \int d^3\xi  \,\bigg\{ \left[ \left(\ds\frac{1}{Z_1} -
1\right) \mathcal{F}_{z\theta} + \ds\frac{K^3}{Z_1V} +
(\mathcal{F}_{tz}-1)\left( \ds\frac{\mu}{Z_1} - \ds\frac{K^1}{V}
\right) \right] \\\\
- \bigg[ \ds\frac{1}{V^2Z_1^2} \big[(K^3  -  V
(\mu(1-\mathcal{F}_{tz})- \mathcal{F}_{z\theta}) )^2
+ V Z_1Z_2 (1-\mathcal{F}_{tz})(2-Z_3(1-\mathcal{F}_{tz})) \big]
\bigg]^{1/2}\bigg\}  \,,
\end{multline}
where $2\pi\alpha' F \equiv \mathcal{F} = \mathcal{F}_{tz} dt\wedge
dz + \mathcal{F}_{z\theta} dz\wedge d\theta$ is the world-volume
gauge field of the D2 brane. Our goal is to semi-classically
quantize BPS fluctuations around certain supertube configurations,
and compute their entropy. Supersymmetry requires that these
fluctuations be independent of $t$ and $z$, and that
${\mathcal{F}_{tz}=1}$.

All the fluctuations of the supertube lead to similar values for the
entropy, but for the purpose of illustrating entropy enhancement it
is best to focus on the fluctuations in the four torus directions:
\begin{equation}
x_i \rightarrow x_i + \eta_{i} (t,\theta) \, \qquad i=1\dots 4.
\end{equation}
Since the BPS modes are independent of $z$, it is convenient to work
with a Lagrangian density that has already been integrated over the
$z$ direction, which gives the conjugate momenta for the excitations
$\eta_i$:
\begin{equation}
\Pi_{i} =  \left( {\partial \over \partial \dot\eta_i}
\int_{0}^{2\pi L_z} \mathcal{L}_{WZ}+\mathcal{L}_{DBI}
\right)_{\dot{\eta}_{i=0},~ \mathcal{F}_{tz}=1} = 2\pi L_z T_{D2}\,
\eta'_{i} \,, \label{CanonMom}
\end{equation}
where $\dot{\eta}_{i} \equiv \frac{\partial\eta_i}{\partial t}$ and
$\eta'_{i} \equiv \frac{\partial\eta_i}{\partial \theta}$.  To
semi-classically quantize the BPS oscillations we impose the
canonical commutation relations:
\begin{equation}
\left[ \eta_j(t,\theta),~ \Pi_{k} (t,\theta') \right] = i
\delta_{jk}\delta(\theta-\theta')  \,. \label{CCRs}
\end{equation}
A supertube with dipole charge $n_2$ can be thought of as wrapped
$n_2$ times around the $\theta$ circle. To find the correct mode
expansion it is not enough to focus on the BPS modes alone, even if
one only wants to count the entropy coming from these modes. Both
the BPS and non-BPS modes contribute to the delta-function in
(\ref{CCRs}) and the inclusion of both contributions is essential to
the proper normalization of the modes\footnote{This subtlety is
correctly taken into account in \cite{Palmer:2004gu}, but not in
\cite{Bak:2004kz}.}.  The result is simply an extra factor of $\sqrt
2$ in the coefficient of the BPS mode expansion compared to the
naive expansion that neglects non-BPS modes:
\begin{equation}
\eta_i =   \eta_i^{\rm~BPS} + \eta_i^{\rm~nonBPS} =
\ds\frac{1}{\sqrt{8\pi^2T_{D2}L_z}} \ds\sum_{k>0} \left[
e^{ik\theta/n_2} \ds\frac{(a^{i}_{k})^{\dagger}}{\sqrt{|k|}} +
\text{h.c.} \right] + \eta_i^{\rm~nonBPS}\,.  \label{etamodes}
\end{equation}
The creation and annihilation operators,  $(a^{i}_{k})^{\dagger}$
and $a_{k}^{i}$, for the modes in the $k^{\rm th}$ harmonic satisfy
canonical commutation relations:
\begin{equation}
[a^{i}_{k}, (a^j_{k'})^{\dagger}] = \delta^{ij}\delta_{k,k'} \,.
\end{equation}
The D0 and F1 quantized charges of the supertube are:
\begin{equation}
Q_{1}^{} =  \frac{T_{D2}}{T_{D0}}  \ds\int_{0}^{2\pi L_z}dz
\ds\int_{0}^{2\pi n_2}d\theta \, \mathcal{F}_{z \theta} = 4\pi^2
\ds\frac{T_{D2}}{T_{D0}} L_z n_2 \mathcal{F}_{z\theta} \label{None}
\end{equation}
\begin{equation}
Q_{3}^{} =  \frac{T_{D2}}{T_{F1}} \ds\int_{0}^{2\pi n_2}d\theta
\left[ -\ds\frac{K^1}{V} + \ds\frac{1}{\mathcal{F}_{z\theta} +
V^{-1}K^3} \left( \ds\frac{Z_2}{V} + (\eta')^2 \right) \right]
 \label{Nthree}
\end{equation}
Substituting (\ref{etamodes}) into (\ref{Nthree})  and rearranging
using (\ref{None})  leads to:
\begin{multline}
\ds\sum_{i=1}^{4}\ds\sum_{k>0} k (a^i_k)^{\dagger} a^i_k = L_z
T_{D2} \int_{0}^{2\pi n_2} d\theta \int_{0}^{2\pi n_2} d\theta' \
\ds\sum_{i=1}^{4} \  \eta'_i\eta'_i \\ = \left[ Q_{1}^{} + 2\pi
T_{F1}L_z n_2 \ds\frac{K^3}{V}\right]\left[ Q_{3}^{} +
\ds\frac{2\pi T_{D2}}{T_{F1}} n_2 \ds\frac{K^1}{V}\right]  - 4\pi^2
T_{D2}L_z n_2^2 \ds\frac{Z_2}{V} \,, \label{EnResult}
\end{multline}
where the integrals over $\theta$ and $\theta'$ are precisely those
appearing in each of (\ref{None}) and (\ref{Nthree}). This is the
result we have been seeking.   The left hand side of
(\ref{EnResult}) can be thought of as the total energy $L_0$ of a
set of four harmonic oscillators in $1+1$ dimensions. For large
$L_0$, the entropy coming from the different ways of distributing
this energy to various modes of these oscillators is  given by the
Cardy formula:
\be S = 2 \pi \sqrt{cL_0 \over 6}. \ee
Since we count BPS excitations, there will be also 4 fermionic
degrees of freedom, and the central charge associated to the torus
oscillations will be
 $c= 4 + 2 =6$, giving the entropy:
\begin{equation}
S = 2\pi \sqrt{\left[ Q_{1}^{} + n_2 \ds\frac{K^3}{V}\right]\left[
Q_{3}^{} + n_2 \ds\frac{K^1}{V}\right]  - n_2^2 \ds\frac{Z_2}{V}}
\, = 2 \pi \, \sqrt{ Q_{1}^{eff} \, Q_{3}^{eff}  ~-~  \, n_2^2
\ds\frac{Z_2}{V}}\,, \label{NiceResult}
\end{equation}
where to render the equations simple we have chosen a system of
units in which $2\pi T_{F1}L_z = L_z/\alpha' = 1$ and $ 2\pi T_{D2}
/T_{F1}  = (g_s\sqrt{\alpha'})^{-1}=1$. We will use this convention
throughout this letter. Equation (\ref{EnResult}) has two important
consequences.   First, for a supertube with a given set of BPS
modes, this equation is nothing but a ``radius formula'' that
determines its size by fixing, in the spatial base, the location of
the $U(1)$ fiber that  it wraps. When the supertube is maximally
spinning, and has no BPS modes, this equation simply becomes the
radius formula of the maximally spinning supertube \cite{bigpaper}.
The second result is that this formula also determines the capacity
of the supertube to store entropy:  In flat space, this capacity is
determined by the asymptotic charges, $Q_{1}^{} $ and
$Q_{3}^{}$, whereas, in a more general background, the capacity to
store entropy is determined by  $Q_{1}^{eff} $ and $Q_{3}^{eff}$. In
certain backgrounds, the latter can be made much  larger than the
former and so a supertube of  given asymptotic charges can have a
lot more modes and thus store a lot more entropy by the simple
expedient of migrating to a location where the effective charges are
very large.  We will discuss this further below.

Clearly, for bubbling backgrounds, and even for black ring
backgrounds, the right hand side of (\ref{EnResult}) can diverge,
and one naively gets an infinite value for the entropy.
Nevertheless, as we mentioned in the introduction, this calculation
is done in the approximation that the supertube does not back-react
on the background, and taking this back-reaction into account will
modify this naive conclusion.

For a supertube that is not along the GH fiber, equation
(\ref{NiceResult}) is still correct, except that the $ Q_{I}^{eff} $
are no longer given by (\ref{QIeff}) but by:
\begin{equation}
Q_{1}^{eff} ~\equiv~  Q_{1}^{}  ~+~ n_2 \, \tilde \zeta^{(1)} \,,
\qquad Q_{3}^{eff} ~\equiv~  Q_{3}^{} ~+~ n_2 \, \tilde\zeta^{(2)}
\,.
\end{equation}
where $\tilde \zeta^{(I)}$ are the pull-backs onto the supertube of
the spacetime one-forms  $\zeta^{(I)}$ defined in
(\ref{GHdipolepotentials}).

We have also explicitly calculated the supertube entropy in a
general three-charge black-ring background, where the supertube
oscillates both in the $T^4$, and in two of the transverse $\IR^4$
directions. The result is identical to (\ref{NiceResult}), except
that now there are six possible bosonic modes (and thus after we
include the corresponding fermions the central charge of the system
is $c=9$). The explicit answer for the entropy\footnote{Using the
conventions of \cite{Bena:2007kg}.} is:
\begin{equation}
S = 2\pi \sqrt{c L_0\over 6} = 2\pi\ds\sqrt{\frac{3}{2}}
\sqrt{\left[ (Q_{1}^{} -
 2 n_2 q_3(1+y)\right]\left[Q_{3}^{} -
 2 n_2 q_1(1+y)\right] -  n_2^2 Z_2
R^2\ds\frac{(y^2-1)}{(x-y)^2}} \,, \label{BRingEnt}
\end{equation}
Based on this result, we expect that upon including the four bosonic
shape modes in the transverse space, as well as the fermionic
counterparts of all the eight bosonic modes, the central charge $c$
should jump from 6 to 12, and equation (\ref{NiceResult}) to be
modified accordingly. We have also explicitly  computed  the entropy
coming from arbitrary shape modes, and the formulas do display
entropy enhancement (they diverge near $y=-\infty$ for the black
ring).  However, the complete expressions are rather unilluminating,
and we leave their study for later investigation \cite{bigpaper}.
Our calculation agrees with the entropy of supertubes in flat
space-time, computed using similar methods in   \cite{Palmer:2004gu,
Bak:2004kz}, and using different methods in \cite{Rychkov:2005ji}.

It is also possible to compute the angular momentum of a supertube
that has a very large number of BPS modes turned on. From the
$T_{0i}$ components of the energy momentum tensor we find
\begin{equation}
J^{ij}= {1 \over 2 \pi } \int_0^{2 \pi n_2} d \theta (\eta_i \Pi_j -
\eta_j \Pi_i)
\end{equation}
and the angular momentum of the tube along the GH fiber is
\begin{equation}
J = {Q_1^{} Q_3^{} \over n_2} - {Q_{1}^{eff} \, Q_{3}^{eff}
\over n_2} + n_2 \ds\frac{Z_2}{V}. \label{Jsuper}
\end{equation}
From this identity we may simply re-write (\ref{NiceResult}) as
\begin{equation}
S ~=~   2 \pi \,\sqrt{ Q_{1}^{eff} \, Q_{3}^{eff}  ~-~  \, n_2^2
\ds\frac{Z_2}{V}}  ~=~    2 \pi \, \sqrt{Q_1^{} Q_3^{} - n_2
J}\,. \label{EntEqns}
\end{equation}
Hence, in a certain sense, (\ref{NiceResult}) is the same as the
entropy formula for a supertube in empty space and it naively
appears that entropy enhancement has gone away.  It has not. The
important point is that (\ref{Jsuper}) implies that it is possible
for $J$ to become extremely large and negative as the number of BPS modes on 
the tube increases\footnote{This is not unexpected: As in flat space, every BPS 
mode on the supertube takes away one quantum of angular momentum of the tube.}. 
In flat space, $|J|$ is limited by $|Q_1^{} Q_3^{} |$ but  in a general
background our Born-Infeld analysis (equations (\ref{EnResult}) and (\ref{Jsuper})) 
imply that the upper bound is the same but there is no lower bound.

From the supergravity perspective, the limits on $J$  usually emerge
from requiring that there are no CTC's near the supertube.  This is
a local condition set by the local behavior of the metric, and
particularly by the $Z_I$, near the supertube.  Although we do not have the explicit solution, our analysis suggests that the lower limit of the 
angular momentum of the supertube is controlled by $Q_{1}^{eff}$ and $Q_{3}^{eff}$ as
opposed to $Q_1^{}$ and $Q_3^{}$. Thus entropy enhancement can
occur if the supertube moves to a region where $Q_{1}^{eff}$ and
$Q_{3}^{eff}$ are extremely large and then a vast number of modes
can be supported on a supertube (of fixed $Q_1^{}$ and $Q_3^{}$)
by making $J$ large and negative. We therefore expect the corresponding supergravity 
solution to be CTC-free provided that $|n_2 J| <  Q_{1}^{eff}
Q_{3}^{eff}$.

One should thus think of a supertube of given $n_2$, $Q_1^{}$
and $Q_3^{}$ as being able to store a certain number of modes
before it over-spins.
The ``storage capacity'' of the supertube is
determined by the local conditions around the supertube and,
specifically, by $n_2$, $Q_1^{eff}$ and $Q_3^{eff}$.  Magnetic
dipole interactions, like those evident in bubbling backgrounds, can
thus greatly modify the capacity of a given supertube to store
entropy.
 
\section{Entropy Enhancement - the Proposal}

As we have seen,  the entropy of a supertube, and hence the entropy
of a fluctuating geometry,  depends upon the {\it local} effective
charges and not upon the asymptotic charges measured at infinity. In the
derivation of (\ref{EnResult}) we started with a maximally spinning,
round supertube with zero entropy and perturbed around it.  For the
maximally spinning tube, the equilibrium position is determined by
the vanishing of the right-hand side.  Upon adding wiggles to the
tube, the right hand side no longer vanishes and the imperfect
cancelation is responsible for the entropy.

It is interesting to ask how much entropy can equation
(\ref{EnResult}) accommodate. The answer is not so simple. At first
glance one might say that the both terms in the right hand side of
(\ref{EnResult}) can be divergent, and hence the entropy of the
fluctuating tube is infinite. Nevertheless, one can see that the
leading order divergent terms in $Q_{1}^{eff} \, Q_{3}^{eff}$ and in
$n_2^2  Z_2/V$ come entirely from bulk supergravity fields, and
exactly cancel, both for the supertube in GH background and for the
supertube near a black ring (\ref{BRingEnt}).

It is likely that this partial cancelation is an artefact of the
extremely symmetric form of the solution, and that in a more general
solution such cancellation may not take place. In particular, both
$Q_1^{eff}$ and $Q_3^{eff}$ are integrals of ``effective charge''
densities on the supertube world-volume, and the right hand side of
equation (\ref{EnResult}) should be written as
\begin{equation}
Q_{1}^{eff} \, Q_{3}^{eff} - n_2^2 \ds\frac{Z_2}{V} = \int{
\rho_{1}^{eff}}  d \theta \, \int{ \rho_{3}^{eff}  d \theta  } -
\int  \rho_{1}^{eff} \, \rho_{3}^{eff}  d \theta \label{Stubeent22}
\end{equation}
If this generalized formula is correct, certain density and shape
modes will disturb the balance between the product of integrals and
the integral of the product, and the leading behavior of the entropy
will still be of the order
\begin{equation}
S  ~\sim~ \sqrt{ Q_{1}^{eff} \, Q_{2}^{eff}  }   \,.
\label{Stubeent2}
\end{equation}

Regardless of this, the next-to-leading divergent terms in
(\ref{NiceResult}) are a combination of supertube world-volume terms
and bulk supergravity fields. In a scaling solution, or when the
tube is close to the black ring, these terms can diverge, giving
naively an infinite entropy. As we discussed above, we expect the
back-reaction of the supertubes to render this entropy finite.

The idea of {\it entropy enhancement} is that one can find
backgrounds in which the effective charges of a two-charge supertube
can be made far larger than the asymptotic charges of the solution,
and that, in the right circumstances, the oscillations of this
humble supertube could give rise to an entropy that grows with the
asymptotic charges much faster than $\sqrt{Q^2}$ (as typical for
supertubes), and might even grow as fast as $\sqrt{Q^{3}}$, as
typical for black holes in five dimensions.

To achieve such a vast enhancement requires a very strong magnetic
dipole-dipole interaction and this means that multiple magnetic
fluxes must be present in the solution.  It is {\it not} sufficient
to have a large red-shift:  BMPV black holes have infinitely long
throats and arbitrarily large red-shifts but have no magnetic dipole
moments to enhance the effective charges and thus increase the
entropy that may be stored on a given supertube.

Hence, the obvious places to obtain entropy enhancement are
solutions with large dipole magnetic fields, such as black ring or
bubbling microstate solutions. Since we are focussing on trying to
obtain the entropy of black holes from horizonless configurations,
we will   focus on the latter. These bubbling solutions are
constructed using an ambi-polar base GH metric, and near the
``critical surfaces,'' where $V$ vanishes,  the term $ {K^I \over V}
$ in the effective charge diverges.  It is therefore natural to
expect entropy enhancement for supertubes that localize near the
critical ($V=0$) surfaces.

We also believe that placing supertubes in deep scaling solutions
\cite{Bena:2006kb,Bena:2007qc,Denef:2007vg} will prove to be an
equally crucial ingredient. Indeed, as we will see in the next
section, in a deep microstate geometry the $K^I$ at the location of
the tube can also become large, and hence there will be a double
enhancement of the effective charge, both because of the vanishing
$V $ in the denominator and because of the very large $K^I$ in the
numerator. There is another obvious reason for this: It is only the
scaling microstate geometries that have the same quantum numbers as
black holes with macroscopic horizons.

This must mean that the simple entropy enhancement one gets from the
presence of critical surfaces is not sufficient for matching the
black hole entropy. The fundamental reason for this may well be the
following: Even if the round supertube can be brought very close to
the $V=0$ surface, once the supertube starts oscillating it will
necessarily sample the region around this surface, and the charge
enhancement will correspond to the average $Q_I^{eff}$ in that
region. For this to be very large the entire region where the
supertube oscillates must have a very significant charge
enhancement. The only such region in a horizonless solution is the
bottom of a deep or scaling throat, where the average of the $K^I$
is indeed very large.

All the issues we have raised here have to do with the details of
the entropy enhancement mechanism, and involve some very long and
complex calculations that we intend to pursue in future work. We
believe their  clarification is very important, as it will shed
light on how the entropy of black holes can be realized at the level
of horizonless configurations.

Our goals in this letter are rather more modest. We have shown via a
Born-Infeld probe calculation that the entropy of supertubes is
given by their effective charges, and not by their brane charges,
and that these effective charges can be very large. However, because
the supertube has been treated as a probe in our calculations, it is
logically possible that, once we take into account its
back-reaction, the bubble equations may forbid the supertube to get
suitably close to the $V=0$ surfaces, and to have a suitable entropy
enhancement.

In principle this is rather unlikely, as we know that in all the
examples studied to date, the solutions of the Born-Infeld action of
supertubes always correspond to configurations that are smooth and
regular in  supergravity \cite{bigpaper}. However, settling the
issue completely is not possible before constructing the full
supergravity solutions corresponding to wiggly supertubes. Hence, in
the remainder of this letter we will show that at least for the
maximally-spinning supertubes, their effective charges in deep
scaling solutions can lead to a black-hole-like enhanced entropy.

\section{Supertubes in scaling microstate geometries}

To find bubbling solutions that contain supertubes with enhanced
charges one could look for solutions of the bubble or integrability
equations \cite{Bena:2005va,Berglund:2005vb,Bates:2003vx}
\begin{equation}
 \sum_{{\scriptstyle j=1} \atop {\scriptstyle j \ne i}}^N \,
\, {\Gamma_{ij}   \over r_{ij} } ~=~  2\, \big(\epsilon_0 \, m_i -
m_0 \, q_i \big) ~+~ \sum_{I=1}^3 \big(\ell^I_i \kappa^I_0 -
\ell^I_0 \, k^I_i \big) ~\,,~~~~~~r_{ij} \equiv |\vec y^{(i)} -\vec
y^{(j)}| \label{BubbleEqns}
\end{equation}
that describe scaling solutions where some of the centers are GH
points, and the other centers are supertubes. However, it is more
convenient to construct such solutions by spectrally flowing
multi-center GH solutions, which have been studied much more. The
parameters of the equations are then:
\begin{equation}
\Pi_{ij}^{(I)} ~\equiv~ \bigg(\frac{k_j^I}{q_j} -\ \frac{k_i^I}{q_i}
\bigg) \,, \qquad\qquad \Gamma_{ij} = q_i \, q_j
\Pi_{ij}^{(1)}\Pi_{ij}^{(2)}\Pi_{ij}^{(3)} \,. \label{PiGamdefn}
\end{equation}
One obtains a scaling solution when a subset, $\cS$, of the GH
points approach one another arbitrarily closely, that is, $r_{ij}
\to 0$ for $i,j \in \cS$.  In terms of the physical geometry, these
points are remaining at a fixed distance from each other, but are
descending a long $AdS$ throat that, in the intermediate region,
looks almost identical to the throat of a black hole or black ring
(depending upon the total GH charge in $\cS$). In particular, in the
intermediate regime, one has $Z_I \sim {\hat Q_I \over 4 \, r}$,
where we have taken $\cS$ to be centered at $r =0$ and the $\hat
Q_I$ are the electric charges associated with $\cS$. Similarly, if
$\cS$ has a non-zero total GH charge of $\hat q_0$, then one has $V
\sim { \hat q_0 \over r}$.  More precisely:
\begin{equation}
Z_I\,V ~=~ l_0^IV ~+~ \varepsilon_0 \, (L_I -  \ell^I_0 )  ~-~
\coeff{1}{4} \, C_{IJK} \, \sum_{i, j=1}^N \, \Pi^{(J)}_{ij} \,
\Pi^{(K)}_{ij} \,    { q_i \, q_j \over  r_i\, r_j}    \,.
\label{ZIVexp}
\end{equation}
Suppose that we perform a spectral flow so that some point, $p \in
\cS$, becomes a supertube. Let $\widetilde V_p$ be the value of
$\widetilde V$ at $p$.  Then, from (\ref{tildefns}) and
(\ref{QIeff}), the effective charges of this supertube are dominated
by terms from interactions with the magnetic fluxes in the throat:
\begin{equation}
Q_I^{eff} ~\sim~ - 2\, q_p \,  \widetilde V_p^{-1} \, C_{IJK} \,
\sum_{j\in \cS\,, \, j \ne p} \, \Pi^{(J)}_{jp} \, \Pi^{(K)}_{jp} \,
{ q_j \over  r_{jp}}    \,. \label{QIeffsim1}
\end{equation}
However, observe that $ \tilde q_j =  (k_p^2)^{-1} q_p q_j
\Pi^{(2)}_{jp}$ and so
\begin{equation}
 q_p^{-1} \widetilde V_p ~\sim~  (k_p^2)^{-1}   \sum_{j\in \cS\,, \, j \ne p}^N \,
  { q_j  \Pi^{(2)}_{jp} \over  r_{jp}}    \,.
\label{Vpsim}
\end{equation}
Therefore the numerator and denominator of (\ref{QIeffsim1}) have
the same naive scaling behavior as $r_{jp} \to 0$ and so, in
general, $Q_I^{eff}$ will attain a finite limit that  only depends
upon the $q_j, k_j^I$ for $j \in \cS$.  Indeed, the finite limit of
$Q_I^{eff}$ scales as the square of the $k$'s for large $k_j^I$
parameters.  This is no different from the typical values of
asymptotic electric charges in bubbled geometries.

However, since we are in a bubbled  microstate geometry, $V$ and
$\widetilde V$ change sign throughout the bubbled region. In
particular, there are surfaces at the bottom of the throat where
$\widetilde V$ vanishes and there are regions around them where
$\widetilde V$ remains finite and bounded as $r_{ij} \to 0$. Suppose
that we can arrange for the supertube point $p$ to be in such a
region of a scaling throat and at the same time we can arrange that
$Z_I$ still diverges as ${1 \over r}$.  Then, in principle, the
effective charges, of the supertube $Q_I^{eff}$, could become
arbitrarily large.

As mentioned above, we expect the entropy of the system to come from
wiggly supertubes in throats that are neither very deep (to allow
the tubes to wiggle), nor very shallow (to give enhancement). We do
not, as yet, know how to take the back-reaction of the wiggly
supertubes into account, and hence we do not have any supergravity
argument about the length of these throats. However, we can use the
$AdS$-CFT correspondence and the fact that we know what the typical
CFT microstates are, to argue \cite{Bena:2006kb} that the typical
bulk microstates are scaling solutions that have GH size $r_T$ given
by
\begin{equation}
 r_{T} ~\sim~ \overline Q^{~-1/2} ~\sim~ {1 \over \bar k }   \,,
\label{rmin}
\end{equation}
where $\overline Q$ is the charge and $\bar k$ is the typical flux
parameter.

If one takes this $AdS$-CFT result as given, and moreover assumes
that the wiggling supertube remains in  a region of finite
$\widetilde V$ in the vicinity of the $\widetilde V=0$ surface, one
then has:
\begin{equation}
Q_I^{eff} ~\sim~  (\bar k)^3 ~\sim~ \overline Q^{~3/2}
\label{QIeffsim}
\end{equation}
because $\Pi^{(K)}_{jp} \sim \bar k$, and hence the entropy of the
fluctuating supertube (\ref{Stubeent2}) would depend upon the
asymptotic charges as:
\begin{equation}
S  ~\sim~ \sqrt{ Q_{1}^{eff} \, Q_{2}^{eff}  } ~\sim~ \overline
Q^{~3/2}   \,. \label{Stubeent3}
\end{equation}
which is precisely the correct behavior for the entropy of a
classical black hole!
 
These simple arguments indicate that fluctuating supertubes at the
bottom of deep scaling microstate geometries can give rise to a
black-hole-like macroscopic entropy, provided that they oscillate in
a region of bounded $\widetilde V$.

Obviously there is a great deal to be checked in this argument,
particularly about the effect of the back-reaction of the supertube
on its localization near the $\widetilde V =0$ surface. We conclude
this section by demonstrating that at least maximally spinning
tubes, for which we can construct the supergravity solution, have no
problem localizing in a region of finite $\widetilde V$. As the
solution scales, the effective charges diverge, as is needed for
entropy enhancement.

\subsection{An example}

One can construct a very simple deep scaling solution using three
Taub-NUT (GH) centers with charges $q_1,q_2$ and $q_3$, and fluxes
arranged so that the  $|\Gamma_{ij}|$, $i,j =1,2,3$, satisfy the
triangle inequalities. The GH points then arrange themselves
asymptotically as a scaled version of this triangle:
\begin{equation}
r_{ij} ~\to~  \lambda \, \big | \Gamma_{ij} \big|  \,,  \qquad
\lambda \to 0 \,. \label{scaling}
\end{equation}
One can then take a spectral-flow of this solution so that the
second GH point becomes a two-charge supertube. For simplicity, we
will choose $ q_1 \, \Pi^{(2)}_{12}~=~  q_3 \,  \Pi^{(2)}_{23}  $ so
that after the flow the GH charges of the remaining two GH points
will be equal and opposite:
\be \tilde q_1 ~=~ -\tilde q_3  \,. \label{eqlchgs} \ee
%
%
%
For $\widetilde V_p$ to remain finite in the scaling limit,  the
supertube must approach  the plane equidistant from the remaining GH
points.

We have performed a detailed analysis of such solutions and used the
absence of CTC's close to the GH points, in the intermediate throat
and in the asymptotic region to constrain the possible fluxes.  We
have found a number of such solutions that have the desired scaling
properties for $Q_I^{eff}$ and we have performed extensive numerical
analysis to check that there are no regions with CTC's.  In
particular, we checked numerically that the inverse metric
component, $g^{tt}$, is globally negative and thus the metric is
stably causal.  We will simply present one example here.

Consider the asymptotically Taub-NUT solution with:
\begin{equation}
q_1 ~=~ 16 \,, \qquad q_2 ~=~ 96 \,, \qquad q_3 ~=~ -40\,, \qquad
\epsilon_0 ~=~1\,, \qquad Q_0  ~\equiv~q_1 +q_2 +q_3 ~=~72
\label{qexample}
\end{equation}
and
\begin{equation}
k_1^I ~=~ (8, -88, 8) \,, \qquad k_2^I ~=~ (0, 96,0)  \,, \qquad
k_3^I ~=~ (20,64,20)   \,, \label{kexample}
\end{equation}
where $Q_0$ is the KK monopole charge of the solution.
 With these parameters one has the following fluxes:
\begin{equation}
\Pi^{(I)}_{12} ~=~ (-\coeff{1}{2}, \coeff{13}{2}, -\coeff{1}{2}) \,,
\qquad \Pi^{(I)}_{23} ~=~ (-\coeff{1}{2}, -\coeff{13}{5},
-\coeff{1}{2})  \,, \qquad \Pi^{(I)}_{13}  ~=~ (-1 , \coeff{39}{10},
-1) \,, \label{Piexample}
\end{equation}
and
\begin{equation}
\Gamma_{12} ~=~ \Gamma_{23} ~=~  \Gamma_{31} ~=~   2496 \,.
\label{gammaexample}
\end{equation}
In this scaling solution the GH points form an equilateral triangle
and thus, after the spectral flow, the supertube will tend to be
equidistant from the two GH points of equal and opposite charges
(\ref{eqlchgs}), and therefore will approach the surface where
$\widetilde V =0$.

The solution to the bubble equations yields
\begin{equation}
 r_{12} ~=~ \frac{11232\, r_{13}}{11232 + 359 \,r_{13}}  \,, \qquad
  r_{23} ~=~ \frac{11232\, r_{13}}{11232 + 731 \, r_{13} }\,,
\label{exbubblesols}
\end{equation}
which satisfies the triangle inequalities for
 $r_{13} \le {11232 \over \sqrt{262429}} \approx 21.9$.
After spectral flow the value of $\widetilde V $ at the location of
the supertube (point 2) is
\begin{equation}
\widetilde V_2~=~ 1 ~+~ \frac{104}{ r_{12}} ~-~ \frac{104}{ r_{23}}
~=~ - \frac{22}{9}\,, \label{exVtilde}
\end{equation}
independent of $r_{13}$.  In particular, it remains finite and
bounded as the three points scale and the distances between them go
to zero. The effective charges of the supertube are given by
\begin{equation}
Q_1^{eff} ~=~ Q_3^{eff} ~=~ 384 \, \widetilde V_2^{-1} \Big( 1 ~+~
\frac{52}{ r_{12}} ~+~
 \frac{52}{ r_{23}}    \Big) \,,
\label{exVtilde1}
\end{equation}
and scale as  $\lambda^{-1}$ as $\lambda \to 0$ in (\ref{scaling}).
We thus have effective charges that naively scale to arbitrarily
large values.  As described earlier,  we expect this scaling to stop
as the supertubes become more and more wiggly, and we expect the
entropy to come from configurations of intermediate throat depth.

Finally, this configuration has asymptotic electric, and
Kaluza-Klein charges:
\begin{equation}
Q_1 ~=~ 416 \,, \qquad Q_2 ~=~ {608 \over 9} \,, \qquad Q_3 ~=~
416\,, \qquad J_R~=~Q^E_{KK} ~=~   {5824 \over 9}\,, \qquad Q_0~=~
Q^M_{KK} ~=~ 72 \,. \label{Qchgexample}
\end{equation}
and is thus a microstate of a Taub-NUT black hole with a finite
extremality parameter and a macroscopic horizon:
\begin{equation}
{Q_0\, Q_1\, Q_2\, Q_3 - {1 \over 4} \, Q_0^2\, J_R^2 \over Q_0\,
Q_1\, Q_2\, Q_3  }   ~=~
 {27 \over 76}~\approx~ 36\%\,.
\label{extremality}
\end{equation}
%

\section{Conclusions}

The most important result   presented in this letter is that the
entropy of a supertube in a given background is not determined by
its charges, but rather by its ``effective charges,'' which receive
a contribution from the interaction of the magnetic dipole moment of
the tube with the magnetic fluxes in the background. As a result,
one can get very dramatic {\it entropy enhancement} if a supertube
is placed in a suitable background.  We have argued that this
enhancement can give rise to a macroscopic (black-hole-like)
entropy, coming entirely from smooth horizonless configurations.

Three ingredients are needed for this dramatic entropy enhancement:
\begin{itemize}
\item[(i)]  Deep or scaling solutions
\item[(ii)]  Ambi-polar base metrics
\item[(iii)]  BPS fluctuations that localize near the critical ($V=0$)  surfaces of the ambi-polar metrics
\end{itemize}
These are also precisely the ingredients that have emerged from
recent developments in the study of finite-sized black-hole microstates in the
regime of parameters where the gravitational back-reaction of some 
of the branes is negligible.   Indeed, deep scaling ambi-polar configurations are
needed both to get a macroscopic entropy in the ``quiver quantum
mechanics regime'' \cite{Denef:2007vg}, and to get smooth
microstates of black holes with macroscopic horizons
\cite{Bena:2006kb}. Furthermore, the D0 branes that can give a
black-hole-like entropy in a D6-$\overline{\rm D6}$ background
\cite{Denef:2007yt} must localize near the critical surface of the
ambi-polar base, much like the supertubes in our analysis. It would
be fascinating to find a link between the microscopic configurations
constructed in these papers, and those we consider here.

In this letter we have referred to the entropy enhancement mechanism
as a ``proposal'' because a number of the details need to be
carefully checked by careful computation. Most importantly, we have
performed a classical calculation using a brane probe near a
critical  surface. It is important to study the fluctuating, or
wiggling, supertubes in the full supergravity theory and determine
how the back-reaction of the fluctuations modifies the picture
presented here.   One important issue is whether fluctuating
supertubes can still remain in the region close to the critical
surface with $V$ finite and bounded. Another is to understand the
interplay between how much a supertube wiggles and how long its
throat can get or how much the supergravity solution it sources can
scale.

While some of the details need to be explored very carefully, we
believe that the mechanism and the approach given in this paper may
well provide the key to understanding how fluctuating microstate
geometries can provide a semi-classical description of black-hole
entropy in the regime of parameters where the black hole exists.

\bigskip
\leftline{\bf Acknowledgements}
\smallskip

We would like to thank Samir Mathur, for interesting discussions. NB
and NPW are grateful to the SPhT, CEA-Saclay for kind hospitality
while this work was completed. The work of NB and NPW was supported
in part by DOE grant DE-FG03-84ER-40168. The work of IB and CR was
supported in part by the DSM CEA-Saclay, by the ANR grant
BLAN06-3-137168, and by the Marie Curie IRG String-QCD-BH. The work
of NB was also supported by the John Stauffer Fellowship from USC
and the Dean Joan M. Schaefer Research Scholarship.



\begin{thebibliography}{99}


\bibitem{review}
S.~D.~Mathur,
  ``The fuzzball proposal for black holes: An elementary review,''
    Fortsch.\ Phys.\  {\bf 53}, 793 (2005)  [arXiv:hep-th/0502050].

 \bibitem{Bena:2007kg}
  I.~Bena and N.~P.~Warner,
``Black holes, black rings and their microstates,''
  arXiv:hep-th/0701216.

\bibitem{kostas}

  K.~Skenderis and M.~Taylor,
  ``The fuzzball proposal for black holes,''
  arXiv:0804.0552 [hep-th].

\bibitem{Bena:2006kb}
  I.~Bena, C.~W.~Wang and N.~P.~Warner,
``Mergers and typical black hole microstates,''
  JHEP {\bf 0611}, 042 (2006)
  [arXiv:hep-th/0608217].

\bibitem{Bena:2007qc}
  I.~Bena, C.~W.~Wang and N.~P.~Warner,
``Plumbing the Abyss: Black Ring Microstates,''
  arXiv:0706.3786 [hep-th].


\bibitem{Bates:2003vx}
  B.~Bates and F.~Denef,
  ``Exact solutions for supersymmetric stationary black hole composites,''
  arXiv:hep-th/0304094.



\bibitem{Mateos:2001qs}
  D.~Mateos and P.~K.~Townsend,
  ``Supertubes,''
  Phys.\ Rev.\ Lett.\  {\bf 87}, 011602 (2001)
  [arXiv:hep-th/0103030].

\bibitem{Lunin:2001jy}
  O.~Lunin and S.~D.~Mathur,
``AdS/CFT duality and the black hole information paradox,''
  Nucl.\ Phys.\  B {\bf 623}, 342 (2002)
  [arXiv:hep-th/0109154].

\bibitem{Lunin:2002iz}
 O.~Lunin, J.~M.~Maldacena and L.~Maoz,
  ``Gravity solutions for the D1-D5 system with angular momentum,'' [arXiv:hep-th/0212210].

\bibitem{Palmer:2004gu}
  B.~C.~Palmer and D.~Marolf,
``Counting supertubes,''
  JHEP {\bf 0406}, 028 (2004)
  [arXiv:hep-th/0403025].

\bibitem{Bak:2004kz}
  D.~Bak, Y.~Hyakutake, S.~Kim and N.~Ohta,
  ``A geometric look on the microstates of supertubes,''
  Nucl.\ Phys.\  B {\bf 712}, 115 (2005)
  [arXiv:hep-th/0407253].

\bibitem{Rychkov:2005ji}
  V.~S.~Rychkov,
  ``D1-D5 black hole microstate counting from supergravity,''
  JHEP {\bf 0601}, 063 (2006)
  [arXiv:hep-th/0512053].

\bibitem{Flowpaper}
I.~Bena, N.~Bobev and N.~P.~Warner,
  ``Spectral Flow, and the Spectrum of Multi-Center Solutions,''
  arXiv:0803.1203 [hep-th].

\bibitem{Giusto:2004kj}
  S.~Giusto and S.~D.~Mathur,
``Geometry of D1-D5-P bound states,''
  Nucl.\ Phys.\  B {\bf 729}, 203 (2005)
  [arXiv:hep-th/0409067].

\bibitem{Ford:2006yb}
  J.~Ford, S.~Giusto and A.~Saxena,
``A class of BPS time-dependent three-charge microstates from
spectral flow,''
  Nucl.\ Phys.\  B {\bf 790}, 258 (2008)
  [arXiv:hep-th/0612227].


\bibitem{Bena:2005va}
  I.~Bena and N.~P.~Warner,
``Bubbling supertubes and foaming black holes,''
  Phys.\ Rev.\  D {\bf 74}, 066001 (2006)
  [arXiv:hep-th/0505166].

\bibitem{Berglund:2005vb}
  P.~Berglund, E.~G.~Gimon and T.~S.~Levi,
``Supergravity microstates for BPS black holes and black rings,''
  JHEP {\bf 0606}, 007 (2006)
  [arXiv:hep-th/0505167].

\bibitem{Denef:2007yt}
  F.~Denef, D.~Gaiotto, A.~Strominger, D.~Van den Bleeken and X.~Yin,
  ``Black hole deconstruction,''
  arXiv:hep-th/0703252; E.~G.~Gimon and T.~S.~Levi,
  ``Black Ring Deconstruction,''
  arXiv:0706.3394 [hep-th].

\bibitem{Denef:2007vg}
  F.~Denef and G.~W.~Moore,
``Split states, entropy enigmas, holes and halos,''
  arXiv:hep-th/0702146.

\bibitem{bigpaper} I. Bena, N. Bobev, C. Ruef and N. P. Warner, ``Supertubes in Bubbling Backgrounds: Born-Infeld Meets Supergravity", \textit{to appear.}

\bibitem{onering}
  I.~Bena and N.~P.~Warner,
  ``One ring to rule them all ... and in the darkness bind them?,''
  Adv.\ Theor.\ Math.\ Phys.\  {\bf 9}, 667 (2005)
  [arXiv:hep-th/0408106].
\end{thebibliography}
\end{document}